\documentstyle[preprint,revtex]{aps}
\begin{document}
\thispagestyle{empty}

\renewcommand{\small}{\normalsize} 

\font\fortssbx=cmssbx10 scaled \magstep2
\hbox to \hsize{
\hskip.5in \raise.1in\hbox{\fortssbx University of Wisconsin - Madison}
\hfill\vtop{\hbox{\bf MAD/PH/757}
            \hbox{\bf NUHEP-TH-93-11}
            \hbox{\bf FERMILAB-PUB-93/092-T}
            \hbox{May 1993}} }
\vspace{.25in}
 \begin{title}
{\bf Finding the Leptonic $WW$ Decay Mode\\
 of a Heavy Higgs Boson\\
at Hadron Supercolliders}
\end{title}
\author{V.~Barger,$^1$ \ Kingman Cheung,$^2$ \ T.~Han,$^3$ \ D.~Zeppenfeld$^1$}
\begin{instit}
$^1$Department of Physics, University of Wisconsin, Madison, WI 53706\\
$^2$Department of Physics \& Astronomy, Northwestern University, Evanston
IL 60208\\
$^3$Fermi National Accelerator Laboratory, P.\ O.\ Box 500, Batavia, IL 60510
\end{instit}
\begin{abstract}
\begin{center} {ABSTRACT} \end{center}
We reanalyze the extraction of the heavy Higgs boson signal $H\rightarrow
W^+W^-\rightarrow \bar\ell\nu,\ell\bar\nu$ $(\ell=e\hbox{ or }\mu)$
{}from the Standard Model background at hadron supercolliders, taking into
account revised estimates of the top quark background. With new
acceptance criteria the detection of the signal remains viable. Requiring
a forward jet-tag, a central jet-veto, and a large relative transverse
momentum of the two charged leptons yields $S/\sqrt B>6$ for one year of
running at the SSC or LHC.
\end{abstract}

\newpage
One of the most important physics issues at hadron supercolliders is the
identification of a heavy Higgs boson in its various production and decay
channels. Single forward jet-tagging (FJT)  has been shown to be very
effective in separating the weak
boson scattering contribution from the gluon fusion process and the QCD
backgrounds in the case of $H\to ZZ\to 4\ell$~\cite{BCHOZ}.
This separation of production
mechanisms is important to fully probe the heavy Higgs sector. Similarly,
it is desirable to independently identify the $H\to ZZ$ and $H\to WW$ decay
modes, in order to test the custodial $SU(2)$ symmetry. Moreover, a neutral
techni-rho $\rho_{TC}^0$ would dominantly decay to $W^+W^-$ rather than $ZZ$.

Recently, the present authors
proposed a method for separation of the leptonic $H\to W^+W^-\to \ell^+\nu
\ell^-\bar\nu$ signal from the large QCD and top-quark pair production
backgrounds~\cite{BCHZ}.
The technique relied on the tagging of a single forward jet to single out the
weak boson scattering process, and the imposition of a central jet-veto (CJV)
to suppress the remaining contribution from top quark pair-production in
association with a QCD jet (denoted by $t\bar t j$), where the
$b$-jets from the top-quark decays populate the central rapidity region.

Subsequent to our work, shower Monte Carlo studies for jet-tagging
were made of the signal and the backgrounds in the $WW$ fusion
channel~\cite{SDC}. Agreement was found with our calculation
except for the amount of $t\bar t j$ background suppression, with the shower
Monte Carlo results finding substantially larger
top backgrounds. This disagreement prompted us to reexamine
our calculation for this channel. We have now identified the source of the
disagreement as a mis-assigned distribution in the output of our computer code:
the energy distribution of the tagging jet in the $t\bar t j$ background was
assigned in the parton center of mass frame rather than the laboratory frame.
Since the two distributions are very different, the $t\bar t j$ background is
found to be higher than originally calculated~\cite{BCHZ}. Therefore
it is necessary for us to
reevaluate the viability of the $H\to W^+W^-$ signal identification above
backgrounds.
Fortunately, we find  a positive conclusion, provided that the relative
transverse momentum of the leptons is required to be large and that
the jet-veto requirement is tightened. Results of our revised
analysis are given below. Apart from the correction in
the computer code the analysis closely parallels that used by us
previously~\cite{BCHZ}.

The leptons arising from the decay of a heavy Higgs boson typically carry
large transverse momenta and they populate the central rapidity region.
As in our original analysis we require
\begin{equation}
\label{lepton}
p_{T\ell}>100$~GeV and $|y_{\ell}|<2\, ,
\end{equation}
throughout. A central feature of our analysis is the tagging of one of the
forward quark-jets arising in the $q_1 q_2\to q_3 q_4 W^+W^-$ signal.
Figure~1 compares the differential $d^2\sigma/dE_jd|\eta_j|$ distribution
in absolute value of pseudorapidity $|\eta_j|$ versus  the energy $E_j$ of the
tagged jet for the signal ($m_H=1$~TeV) and the $t\bar t j$
background.  By selecting a region of high rapidities and substantial tagging
jet energies,
\begin{equation}
\label{tag}
3 < |\eta_j({\rm tag})| < 5\,,  \ \  E_j({\rm tag})>1~{\rm TeV}\,,
\ \ {\rm and} \ \ p_{Tj}({\rm tag})>40~{\rm GeV}\,,
\end{equation}
the backgrounds with QCD jet emission are suppressed relative to the signal.
The benefits of a more stringent jet-tagging requirement will be discussed
later.

The effectiveness of a central jet-veto is demonstrated in Fig.~2 which shows
the pseudorapidity distribution of the second jet (veto candidate).
While the signal events rarely produce a central jet with $p_{Tj}>30$~GeV, the
rate of such jets in the $t\bar tj$ background  is quite large. We
tighten the central jet-veto cut of Ref.~\cite{BCHZ}, and reject events with
\begin{equation}
\label{veto}
p_{Tj}{\rm (veto)} >30~{\rm GeV}, \ \ |\eta_j({\rm veto})|<3,
\end{equation}
which is still above the 25 GeV central-jet threshold used by the  SDC
Collaboration~\cite{SDC}.  Eq.(\ref{veto}) will be the central jet-veto (CJV)
requirements for background suppression. The combined efficiency of
the FJT and CJV for a 1~TeV Higgs-boson signal
is about 40\%, while these cuts reduce the $t \bar t j$
background by about 3 orders of magnitude.

The charged leptons originating from the heavy Higgs-boson
 signal have higher $p_T$ (typically $m_H/4$)
and are more back-to-back than the backgrounds~\cite{BCHP,DGOV,BC}. We find
that the distribution in
\begin{equation}
\Delta p_{T\ell\ell} = | {\bf p}_{T\ell_1}-{\bf p}_{T\ell_2}|,
\end{equation}
which has been considered previously in studying $W^+W^+\to W^+W^+$
scattering~\cite{DGOV},
is an appropriate vehicle to reduce the $t\bar t j$ background
to acceptable levels. The $\Delta p_{T\ell\ell}$
distributions for the signal and the various background processes at
the SSC energy are shown in Fig.~3. By requiring
\begin{equation}
\label{dptll}
\Delta p_{T\ell\ell}>400\;{\rm GeV}
\end{equation}
we obtain a sufficient suppression of $t\bar t j$ events.

We summarize the effects of the acceptance cuts in two tables, for the SSC and
LHC, respectively.
The first line in Tables~I and II gives the cross sections (in fb) for the
1~TeV and 0.6~TeV Higgs-boson cases, the electroweak
transverse $W$ background (estimated with the $m_H=0.1$~TeV SM expectation),
the QCD background, the $t\bar tj$ background (for $m_t=140$ and 180~GeV),  and
the significance $S/\sqrt B$ of the $m_H=1$~TeV signal estimated with
$m_t=140$~GeV and
an integrated luminosity of $10~{\rm fb}^{-1}$ ($100~{\rm fb}^{-1}$) at the
SSC (LHC). Thus a high level of
significance can be achieved, even if uncertainties in background normalization
are folded in.

Beyond the acceptance criteria used above, the kinematic distributions of the
heavy Higgs signal and the backgrounds differ substantially. As examples Fig.~4
gives the energy distribution of the tagged jet at the SSC and the LHC, Fig.~5
shows the distributions in $p_{T\ell}^{\rm max}$ of the lepton with
the maximal transverse
momentum, and Fig.~6 gives the distribution in $m_{\ell j}^{\rm min}$ of the
smaller of the two lepton-tagging jet invariant masses. In all three examples
the distribution of the signal is much flatter than that of the major
backgrounds.

Because the signal and background distributions are quite different in shape,
the positive identification of a Higgs signal is independent of modest
normalization uncertainties in the prediction of the signal and background
cross
sections. A simultaneous fit to the shape of all the available distributions
is the most promising means for an unambiguous extraction of the heavy
Higgs signal. Short of such a  complete analysis, one can
improve the significance of the signal by more stringent acceptance criteria,
at modest cost to the signal rate. The last few lines in the two tables
provide illustrations. Although the large $\Delta p_{T\ell\ell}$ criterion
seems to be most effective, large $p_{T\ell}^{\rm max}$ is useful too.
Cuts on the $m_{\ell j}^{\rm min}$ variable~\cite{DGOV} appear promising as
well, the uncertainties on the energy and direction of the tagging
jet may mitigate its usefulness, however.

Our analysis was largely based on an assumed top mass of 140 GeV.
The suppression of the top
background is easier for heavier $m_t$ because $b$ quarks from the top decay
have higher $p_T$ and the central jet-veto is more effective. This is
illustrated in Tables~I and II by the $m_t=180$~GeV columns: the dominant top
quark background is reduced by a factor 1.5.

In addition to the forward jet-tagging and the central jet-vetoing, the
$\Delta p_{T\ell\ell}>400$~GeV cut is crucial for the top quark background
reduction. This cut is specifically tailored for the case of a 1~TeV Higgs. As
can be seen
{}from the tables, this cut starts to be too severe for Higgs masses around or
below 0.6~TeV. For such Higgs masses a relaxed $\Delta p_{T\ell\ell}$ cut
combined with a more
stringent cut on the tagging jet energy $E_j$(tag) would be desirable, as can
be deduced from the effects of increasing these cuts in Tables~I
and II for the $m_H=600$~GeV case.

We conclude that the prospects for finding the $WW$ leptonic decays of the
heavy Higgs boson at the LHC or SSC remain very good, in spite of the fact that
the $t\bar t j$ background is larger than previously indicated. We have found
improved selection criteria which make an effective background suppression
still
possible. In addition to the forward jet-tag and central jet-veto a substantial
relative $p_T$ of the two leptons must be demanded.

\acknowledgements
We are grateful to K.~Einsweiler,
F.~Paige, L.~Orr, and S.~Willenbrock for helpful discussions.
This research was supported in part by the University of Wisconsin Research
Committee with funds granted by the Wisconsin Alumni Research Foundation,
in part by the U.~S.~Department of Energy under Contract No.~DE-AC02-76ER00881,
and in part by the Texas National Research Laboratory Commission under Grants
No.~RGFY9273, FCFY9212 and FCFY9116.

\newpage             

\setdec 0.00

\begin{table}
\caption{SSC cross section in fb for various acceptance cuts on $W^+W^-jX$
events with leptonic $W$ decays. The tagging requirements ($E_j({\rm
tag})>1$~TeV, $p_{Tj}({\rm tag})>40$~GeV, and $3<|\eta_j({\rm tag})|<5$),
central
jet-veto ($p_{Tj}({\rm veto})>30$~GeV and $|\eta_j({\rm veto})|<3$), and
generic lepton cuts ($p_{T\ell}>100$~GeV, $\Delta p_{T\ell\ell}>400$~GeV, and
$|y_\ell | < 2$) are imposed throughout. In addition results are shown for a
selection of enhanced acceptance cuts. The final column gives the significance
$S/\sqrt B$ for $m_H=1$~TeV, $m_t=140$~GeV, and an integrated luminosity of
10 fb$^{-1}$, corresponding to one year of running at design luminosity.}
\begin{tabular}{lccccccc}
\underline{Further cuts}& \multicolumn{3}{c}{\underline{$m_H$ (TeV)}}&
\underline{QCD}& \multicolumn{2}{c}{\underline{$t\bar tj$}}
& \underline{$S/\sqrt B$}\\
 & 1.0& 0.6& 0.1&& $m_t=140$& $m_t=180$~GeV &  \\
\tableline
no additional & \dec 5.8 & \dec 2.5  & \dec 0.54 & \dec 0.79 & \dec 4.3 & \dec
2.8 & \dec 7.1  \\
$E_j(\rm tag) > 1.5$ TeV & \dec 5.1 & \dec 2.2 & \dec 0.49 & \dec 0.51
& \dec 2.8 & \dec 1.8  & \dec 7.5  \\
$\Delta p_{T\ell\ell}>450$ GeV & \dec 4.8 & \dec 1.4 & \dec 0.42 & \dec 0.59
& \dec 2.5 & \dec 1.5  & \dec 7.4   \\
$p_{T\ell}^{\rm max}>270$ GeV  & \dec 5.0  & \dec 1.6 & \dec 0.46 & \dec 0.59 &
\dec 2.6 & \dec 1.6  & \dec 7.4   \\
$m_{\ell j}^{\rm min}>500$ GeV& \dec 5.2 & \dec 2.2 & \dec 0.44 & \dec 0.50 &
\dec 3.0 & \dec 1.9 & \dec 7.6
\end{tabular}
\end{table}

\newpage

\setdec 0.000
\begin{table}
\caption{LHC cross section in fb for various acceptance cuts on $W^+W^-jX$
events with leptonic $W$ decays. Acceptance cuts are as in Table~1 except for a
relaxed tagging requirement $E_j({\rm tag})>0.8$~TeV.
The final column gives the significance
$S/\sqrt B$ for $m_H=1$~TeV, $m_t=140$~GeV, and an integrated luminosity of
100 fb$^{-1}$, corresponding to one year of running at design luminosity.}
\begin{tabular}{lccccccc}
\underline{Further cuts}& \multicolumn{3}{c}{\underline{$m_H$ (TeV)}}&
\underline{QCD}& \multicolumn{2}{c}{\underline{$t\bar tj$}}
& \underline{$S/\sqrt B$}\\
 & 1.0& 0.6& 0.1&& $m_t=140$& $m_t=180$~GeV&\\
\tableline
no additional & \dec 0.46 & \dec 0.25  & \dec 0.044 & \dec 0.11 & \dec 0.34 &
\dec 0.21 & \dec 5.9 \\
$E_j(\rm tag) > 1.0$ TeV & \dec 0.42 & \dec 0.23 & \dec 0.041 & \dec 0.078
& \dec 0.28 & \dec 0.17  & \dec 6.0 \\
$E_j(\rm tag) > 1.2$ TeV & \dec 0.38 & \dec 0.21 & \dec 0.038 & \dec 0.059
& \dec 0.23 & \dec 0.14  & \dec 6.0 \\
$\Delta p_{T\ell\ell}>450$ GeV & \dec 0.37 & \dec 0.13 & \dec 0.031
& \dec 0.074 & \dec 0.17 & \dec 0.11 & \dec 6.5 \\
$p_{T\ell}^{\rm max}>270$ GeV & \dec 0.37 & \dec 0.15 & \dec 0.034
& \dec 0.067 & \dec 0.18 & \dec 0.11 & \dec 6.3 \\
$m_{\ell j}^{\rm min}>500$ GeV& \dec 0.38 & \dec 0.20 & \dec 0.032
& \dec 0.054 & \dec 0.21 & \dec 0.13 & \dec 6.4
\end{tabular}
\end{table}


\figure{\label{one} $d^2\sigma/dE_jd|\eta_j|$ distributions of the tagged jet
at the SSC from
(a) the $m_H=1$ TeV SM signal, and
(b) the $t \bar tj$ background for $m_t=140$\,GeV. The acceptance cut of
Eq.~(\ref{lepton}) are imposed. }

\figure{\label{two} Pseudorapidity distributions of the second jet
(veto candidate) for  the $t \bar tj$, electroweak $qqWW$
($m_H=0.1$~TeV) backgrounds and the $m_H=1$~TeV SM Higgs boson signal at
the SSC with a  tagging jet requirement of $E_j>1$~TeV. The acceptance
cuts are those of Eqs.~(\ref{lepton}), (\ref{tag}), and (\ref{dptll})
 and $p_{Tj}({\rm veto})>30$~GeV. }

\figure{\label{three} Relative transverse momentum distribution
$\Delta p_{T\ell\ell}$ for the signal and the various background processes
at the SSC, with the acceptance  cuts of Eqs.~(\ref{lepton}),
(\ref{tag}), and (\ref{veto}).  }

\figure{\label{four} Energy distribution (a) at the SSC and (b) at the LHC
of the tagged jet, with the acceptance cuts of Eqs.~(\ref{lepton}),
(\ref{tag}), (\ref{veto}), and (\ref{dptll}).  }

\figure{\label{five} Distribution in transverse momentum $p_{T\ell}^{\rm max}$
of the charged lepton with the maximum $p_T$ in each event.
Acceptance cuts are as in Fig.~\ref{four}. }

\figure{\label{six} Distribution in the smallest invariant mass of a charged
lepton with the tagging jet. Acceptance cuts are as in Fig.~\ref{four}. }

\end{document}